\documentclass{iaus}
\usepackage{graphicx}

\title[From nuclear clusters to halo GCs] 
{From nuclear clusters to halo globulars: Star clusters as basic galactic building blocks}

\author[Richard de Grijs]   
{Richard de Grijs}

\affiliation{Department of Physics \& Astronomy, The University of
Sheffield, Hicks Building, Hounsfield Road, Sheffield S3 7RH, UK}

\pubyear{2006}
\volume{235}  
\jname{Galaxy Evolution across the Hubble Time}
\editors{Fran\c{c}oise Combes and Jan Palou\v{s}, eds.}
\begin{document}

\maketitle

\begin{abstract}
I assess the similarities and differences between the star-formation
modes in quiescent spiral galaxies versus those in violent starburst
regions, including galactic nuclei. As opposed to the quiescent
star-formation mode, current empirical evidence on the star-formation
processes in the extreme, high-pressure environments induced by galaxy
encounters strongly suggests that star {\it cluster} formation is an
important and perhaps even the dominant mode of star formation in such
starburst events. This implies that by using star clusters as unique
diagnostic probes, we can trace a galaxy's most violent star formation
history very well, at least for the past few Gyr. The sizes,
luminosities, and mass estimates of the young massive star clusters
are entirely consistent with what is expected for young Milky Way-type
globular clusters (GCs). Recent evidence lends support to the scenario
that GCs, which were once thought to be the oldest building blocks of
galaxies, are still forming today.
\end{abstract}

\section{Young massive star clusters as proto-globular clusters}

Luminous, massive yet compact star clusters (YMCs; often with masses
$m_{\rm cl} \ge 10^5 {\rm M}_\odot$) are the hallmarks of the most
intense starbursts. YMCs are therefore important as benchmarks of
cluster formation and evolution, and also as important tracers of the
(stellar) initial mass function (IMF) and other physical properties of
starbursts. This is so, because star clusters can extremely well be
represented by single-age, single-metallicity ``simple'' stellar
populations (SSPs). This sets the YMCs formed in violent starburst
events apart from the apparently similar nuclear star clusters, since
the latter are often characterised by stellar population mixes (e.g.,
Walcher et al. 2006). The defining characteristics of YMCs have been
explored in starburst regions in several dozen galaxies, both in
normal spirals and in gravitationally interacting galaxies.

The crucial question remains, however, whether or not at least some of
the YMCs observed in extragalactic starbursts might survive to become
(possibly somewhat more metal-rich) counterparts of the Galactic
globular clusters (GCs) when they reach a similar age. If we could
resolve this issue convincingly, one way or the other, the
implications would be far-reaching for a wide range of astrophysical
questions, including our understanding of how galaxy formation,
assembly and evolution proceeds, and what the process and requisite
conditions are for star (cluster) formation.

The evolution of young clusters depends crucially on their stellar
IMF: if its slope is too shallow, i.e., if the clusters are
significantly deficient in low-mass stars compared to, e.g., the solar
neighbourhood, they will likely disperse within about a Gyr of their
formation (e.g., Gnedin \& Ostriker 1997; Goodwin 1997; Smith \&
Gallagher 2001; Mengel et al. 2002; de Grijs \& Goodwin, in prep.). At
present, there are two principal approaches which one can use to
address the underlying IMFs of extragalactic YMCs.

\subsection{The Cluster Luminosity Function: the case of M82}

In de Grijs et al. (2003a,b) we reported the discovery of an
approximately log-normal cluster luminosity and mass function (CLF,
CMF) for the subpopulation of approximately coeval star clusters at
the intermediate age of $\sim 1$ Gyr in M82's fossil starburst region
``B''.  This provided the first deep CLF (CMF) for a star cluster
population at intermediate age, which thus serves as an important
benchmark for theories of the evolution of star cluster systems [see
also Goudfrooij et al. (2004) for a related important result for NGC
1316, at $\sim 3$ Gyr]. The fact that we considered an approximately
coeval subset of the M82 B cluster population, combined with our use
of the 100 per cent completeness limit as our base line ensures the
robustness of the CMF peak detection. Additional arguments in favour
of this robustness are provided in de Grijs et al. (2005).

Most analyses of the young Large Magellanic Cloud cluster system (with
ages $\le 2 \times 10^9$ yr), starting with the seminal work by Elson
\& Fall (1985), seem to imply that the CLF of YMCs is well described
by a power law. However, in de Grijs \& Anders (2006) we showed that
this power-law behaviour breaks down below $m_{\rm cl} \simeq 10^3
{\rm M}_\odot$, as well as for our youngest LMC cluster age
bins. Although the latter could partially be due to the so-called
``infant mortality'' effect, by which some 60 -- 90 per cent of newly
formed clusters are disrupted in the first $\sim 10$ Myr of their
lifetime (e.g., Bastian et al. 2005), the deviation of the LMC CLFs at
low cluster masses is significant for ages up to at least 1 Gyr,
beyond which our observational completeness limit prevents us from
making definite statements on the CLF shape at low mass (de Grijs \&
Anders 2006).

Nevertheless, for old GC systems with ages $\ge 10^{10}$ yr, the CLF
shape is -- on the other hand -- well established to be roughly
log-normal, and almost universal among local galaxies (barring a
slight metallicity dependence). This type of observational evidence
has led to the popular, but thus far mostly speculative theoretical
prediction that not only a power-law, but {\it any} initial CLF (CMF)
will be rapidly transformed into a log-normal distribution because of
{\it (i)} stellar evolutionary fading of the lowest-luminosity (mass)
objects to below the detection limit; and {\it (ii)} disruption of the
low-mass clusters due both to interactions with the gravitational
field of the host galaxy, and to internal two-body relaxation effects
leading to enhanced cluster evaporation (but see Parmentier \& Gilmore
2006).

From our detailed analysis of the expected evolution of CMFs starting
from initial log-normal and initial power-law distributions (de Grijs
et al. 2005), we conclude that the observed turnover in the M82 B CMF
is inconsistent with a scenario in which the 1 Gyr-old cluster
population originated from an initial power-law mass distribution.
This applies to a large range of ``characteristic'' cluster disruption
time-scales, and is supported by arguments related to the initial
density in M82 B, which would be unphysically high if the present
cluster population were the remains of an initial power-law
distribution (particularly in view of the effects of the cluster
``infant mortality'', which requires large excesses of low-mass
unbound clusters to be present at the earliest times).

In de Grijs et al. (2003c) we showed that the CMFs of YMCs in many
different environments are well approximated by power laws with slopes
$\alpha \simeq -2$. However, except for the intermediate-age cluster
systems in M82 B and NGC 1316 (Goudfrooij et al. 2004), the {\it
expected} turn-over (or peak) mass (based on comparisons with
present-day GC systems and taking evolutionary fading into account) in
most YMC systems observed to date occurs close to or below the
observational detection limit, simply because of their greater
distances and shallower observations. As such, these results are not
necessarily at odds with each other, but merely hindered by
observational selection effects.

\subsection{High-resolution spectroscopy: individual cluster analysis}

With the ever increasing number of large-aperture ground-based
telescopes equipped with state-of-the-art high-resolution
spectrographs and the wealth of observational data provided by the
{\sl Hubble Space Telescope}, we may now finally be getting close to
resolving the potentially far-reaching issue of YMC-to-GC evolution
conclusively. To do so, one needs to obtain {\it (i)} high-resolution
spectroscopy, in order to obtain dynamical mass estimates, and {\it
(ii)} high-resolution imaging to measure their sizes (and
luminosities). As a simple first approach, one could then construct
diagnostic diagrams of YMC mass-to-light ratio vs. age, and compare
the YMC locations in this diagram with SSP models using a variety of
IMF descriptions (cf. Smith \& Gallagher 2001; Mengel et al. 2002;
Bastian et al. 2006). However, such an approach, while instructive,
has serious shortcomings:

{\it (i)} In this simple approach, the data can be described by {\it
both} variations in the IMF slope {\it and} variations in a possible
low-mass cut-off; the models are fundamentally degenerate for these
parameters.

{\it (ii)} While the assumption that these objects are approximately
in virial equilibrium is probably justified at ages greater than a few
$\times 10^7$ yr (at least for the stars dominating the light), the
{\it central} velocity dispersion (as derived from luminosity-weighted
high-resolution spectroscopy) does not necessarily represent a YMC's
total mass. It is now well-established that almost every YMC exhibits
significant mass segregation from very early on, so that the effects
of mass segregation must be taken into account when converting central
velocity dispersions into dynamical mass estimates (see also Lamers et
al. 2006; Fleck et al. 2006; Moll et al. 2006 and in prep.).

{\it (iii)} With the exception of a few studies (e.g., M82-F; Smith \&
Gallagher 2001), the majority of YMCs thus far analysed in this way
have ages around 10 Myr. Around this age, however, red supergiants
(RSGs) appear in realistic stellar populations. Unfortunately, the
model descriptions of the RSG phase differ significantly among the
various leading groups producing theoretical stellar population
synthesis codes (Padova vs. Geneva vs. Yale), and therefore the
uncertainties in the evolutionary tracks are substantial.

{\it (iv)} For nuclear star clusters, one also needs to relax the SSP
approximation, and allow multiple stellar populations to be present in
the mix; this has the potential of matching the observed
high-resolution spectra very well, however (Walcher et al. 2006).

\section{Where does this leave us?}

It may appear that a fair fraction of the $\sim 10$ Myr-old YMCs that
have been analysed thus far may be characterised by unusual IMFs,
since their loci in the diagnostic diagram are far removed from any of
the ``standard'' SSP models (see, e.g., Bastian et al. 2006; Moll et
al. 2006 and in prep.). However, Bastian \& Goodwin (2006) recently
showed that this is most likely an effect of the fact that the
velocity dispersions of these young bjects do not adequately trace
their masses. They are instead strongly affected by the effects of gas
expulsion due to supernova activity and massive stellar winds. In this
respect, it is encouraging to see that the older clusters (i.e., older
than M82-F, a few $\times 10^7$ yr) seem to conform to ``normal''
IMFs; by those ages, the clusters' velocity dispersions seem to
represent the underlying gravitational potential much more closely.

We recently reported the discovery of a extremely massive, but old
($12.4 \pm 3.2$ Gyr) GC in M31, 037-B327, that has all the
characteristics of having been an exemplary YMC at earlier times (Ma
et al. 2006). To have survived for a Hubble time, we conclude that its
stellar IMF cannot have been top-heavy, i.e., characterized by a
low-mass cut-off at $m_\star \ge 1$ M$_\odot$, as sometimes advocated
for current YMCs (e.g., Smith \& Gallagher 2001). Using this
constraint, and a variety of SSP models, we determine a photometric
mass for 037-B327 of $M_{\rm GC} = (3.0 \pm 0.5)\times 10^7$
M$_\odot$, somewhat depending on the SSP models used, the metallicity
and age adopted and the IMF representation. In view of the large
number of free parameters, the uncertainty in our photometric mass
estimate is surprisingly small. This mass, and its relatively small
uncertainties, make this object the most massive star cluster of any
age in the Local Group. As a surviving ``super'' star cluster, this
object is of prime importance for theories aimed at describing massive
star cluster evolution.

\begin{discussion}

\discuss{Emsellem}{Could the ``super'' stellar cluster you studied be
the remnant nucleus of a galaxy that has been stripped?}

\discuss{de Grijs}{Although this cannot be ruled out completely at
present, I suspect that it really is a genuine GC. In addition to the
fact that its SED is very well fit by a simple stellar population, it
has a very small size ($R_{\rm hl} \sim 2.5 \pm 0.2$ pc at the
distance of M31, $m-M = 24.88$ mag), and shows no evidence of an
extended envelope (Ma et al. 2006).}

\discuss{Block}{Does your LMC cluster sample span all the ages
contained in the Searle, Wilkinson \& Bagnuolo cluster classification
types I -- VII?}

\discuss{de Grijs}{Yes, it does. I refer you to Fig. 7 in de Grijs \&
Anders (2006), where we show our full LMC sample in
log(age)--log(mass) space, including the 50 per cent completeness
limit. As you will notice, our LMC sample ranges from the youngest
clusters (returned as 4 Myr because of limitations of our models at
young ages) to the well-known old GCs, with ages $> 10$ Gyr.}

\end{discussion}

\end{document}